 \documentstyle[multicol,aps]{revtex}
\begin{document}
\include{epsf}
\draft

%
%

%
%
\title {\Large \bf 
Mesoscopic molecular ions in Bose-Einstein condensates}

\author{R. C\^{o}t\'{e} $^1$, V.~Kharchenko $^2$, and M.~D.~Lukin$^{2,3}$}

\address{$^1$Physics Department, University of Connecticut, 
2152 Hillside Rd., Storrs, Connecticut 06269-3046\\
$^2$ ITAMP, Harvard-Smithsonian Center for Astrophysics, 60 Garden Street,
Cambridge, MA 02138 \\
$^3$ Physics Department, Harvard University, Cambridge, MA 02138
}
\date{\today}
\maketitle
\begin{abstract}

We study the possible formation of large (mesoscopic) molecular
ions in an ultracold degenerate bosonic gas doped with charged 
particles (ions). 
We show that the polarization potentials produced by the
ionic impurities are capable of capturing hundreds of
atoms into loosely bound states. We describe the spontaneous 
formation of these hollow molecular ions via phonon emission and 
suggest an optical technique for coherent stimulated transitions of free atoms
into a specific bound state.  These results open up new interesting 
possibilities for manipulating tightly confined ensembles.

\end{abstract}


  \begin{multicols}{2}

The experimental realization of Bose-Einstein 
condensation (BEC) in atomic samples \cite{bec} have stimulated 
a new wave of theoretical and experimental studies of degenerate 
systems in the dilute and weakly interacting regime.  
A broad range of techniques from
atomic physics and optics have allowed the accurate 
manipulation of ultracold samples of neutral atoms. 
In particular, various properties of degenerate 
gases have been explored,
such as their excitation modes \cite{excitation}, 
superfluidity phenomena \cite{fluid,eddy}, and controlled 
vortex creation \cite{vortex}. Recent studies are 
probing ultracold atomic systems in which electric 
charges may play an important role \cite{hopping}. 
These include ultracold plasmas \cite{cold-plasma}, 
ultracold Rydberg gases \cite{cold-rydberg}, as well as 
ionization experiments in BEC \cite{arimondo}.

In this Letter, we explore theoretically the behavior of a dilute 
atomic Bose-Einstein condensate doped with ionic impurities. 
We show that the polarization interaction in a condensate can 
lead to the capture of large numbers of atoms into weakly bound states, 
resulting in the rapid formation of mesoscopically large molecular ions. 
We study the spontaneous dynamics of the molecular ion formation 
and show that the degenerate nature of the condensate, as well as the 
properties of collective excitations (phonons), play an important role.
We further describe a coherent optical technique to prepare 
molecular ions in specific states. Beside the fundamental interest
of studying the formation of such large many-body objects,
the effects described here may open up new ways to 
manipulate cold atoms.  In particular, the charged and tightly 
trapped atomic cloud represents a microtrap that could easily be 
manipulated and ``transported'' 
by external fields. Controlled mechanisms
to manipulate tightly confined, strongly interacting atoms 
may also allow for new approaches for quantum information processing
and for studies of other fascinating phenomena such as quantum phase 
transitions. Before proceeding we note an interesting analogy to 
early studies involving charged impurities in superfluid 
helium, where electron bubbles and ion ``snowballs" were predicted and 
observed \cite{helium}.

We consider the situation in which few ions with ultralow kinetic energy  
are introduced into an atomic BEC. This can be realized, e.g., by rapidly
ionizing atoms from the condensate using lasers in a 
process where the ejected
electrons carry essentially all the kinetic energy. This would leave
the BEC doped with few ions, and the BEC-ion system in a non-equilibrium 
state. Alternatively, it could be possible to introduce charged 
impurities via controlled processes involving either a combination of 
ion and atom traps, or using surface traps on semiconductor surfaces. 
As a relaxation process, large numbers of atoms from the condensate
can be captured into loosely bound states of the polarization potentials, 
rapidly forming shells of atoms around ions. 
Such a process occurs spontaneously through collisions of 
condensate atoms in which atoms are stimulated down into the 
molecular ion bound state, and the excess energy is carried 
away by the condensate collective excitations.

We are interested in the limit $T\rightarrow 0$, and 
for simplicity, we consider a homogeneous BEC with
the neutral gas being the parent atom of the doping ion
\cite{note-ion}. 
The ion will polarize a nearby atom (separated by a distance $r$) 
and the two will interact 
via a polarization potential
behaving asymptotically as $-C_{4}/2 r^{4}$, where
$C_{4}$ is the dipole polarizability of the neutral atom.

\begin{figure}[h]
 \centerline{\epsfxsize=3.25in\epsfbox{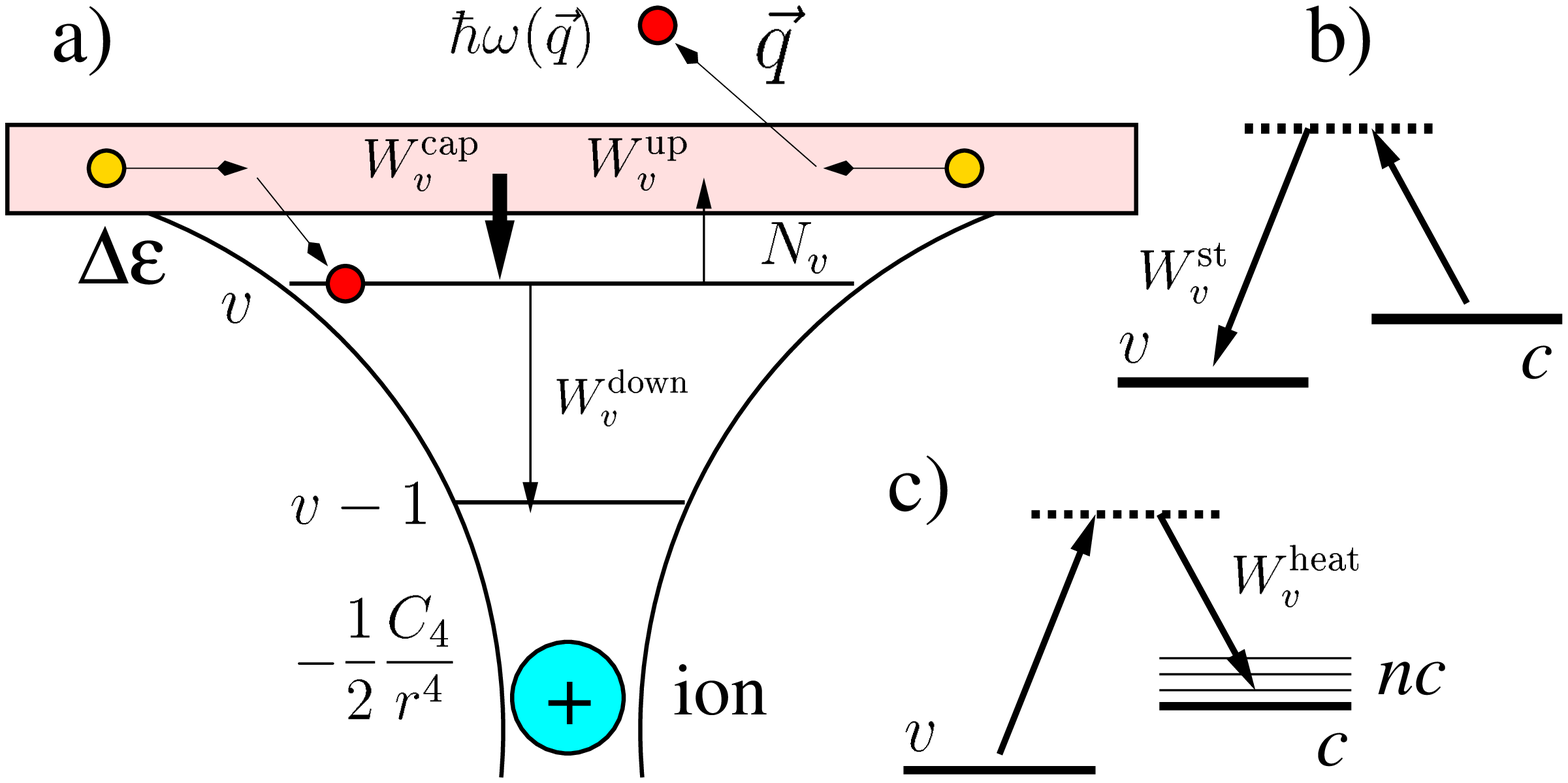}}
\caption{ \protect\narrowtext
          Diagrams of atom capture by an ion. In (a), the spontaneous 
          capture in level $v$ 
          is followed phonon emission (with corresponding rates).
          In (b), the stimulated rate from the condensate $c$ 
          to the bound level $v$, also may produce heating, i.e.
          population of noncondensate atoms $nc$, in (c).}
\label{fig1}
\end{figure}

Under the conditions mentioned above, two types of scattering
processes involving the ion are possible: elastic and super-elastic 
collisions. The energy of the colliding partners 
is not changed in the first type, while kinetic energy 
is released in the second type. 
Elastic scattering of an atom and an ion is often described as 
the contributions from two separate processes \cite{hopping,mcdaniel}:
resonant charge transfer scattering (the electron changes center 
after the collision), and pure elastic scattering (the electron remains
with the same center). According to Wigner's threshold 
laws \cite{wigner}, the rate of all elastic processes vanishes 
as $T\rightarrow 0$. The second type of scattering process, i.e. 
super-elastic scattering, corresponds to 
an inelastic process where one condensate atom is
captured by the ion and where kinetic energy is released via the
emission of phonons. Here, condensate atoms 
are accelerated by the ionic field (see Fig.~\ref{fig1}a), and 
after the collision, one of the neutral atoms is captured by the 
polarization potential, and the kinetic energy released is shared 
by the remaining free condensate atoms and the newly formed molecular 
ion \cite{note-recoil}.
Inelastic collisions with excitations of BEC 
collective modes usually cannot be decribed by simple binary 
collisions, unless in the limit where the binding energy 
of the uppermost bound level of the molecular ion is much 
larger than the chemical potential of the BEC. 
According to Wigner's threshold laws \cite{wigner},
the rate for super-elastic processes tends to a constant
at zero temperature, and this scattering process
will be dominant at $T\rightarrow 0$. 
Contrary to the slow-down of impurities in condensates, where 
phonon radiation is not possible below some critical velocity 
(because of momentum-energy conservation \cite{eddy}), the
capture of atoms by the long range potential created by the
ion corresponds to free-bound
transitions, and does not suffer from this restriction:
phonon emission takes place at any velocity.

We describe the evolution of the number of atoms $N_{v}$
in the bound level $v$ of the polarization potential
by a kinetic equation (see Fig.~\ref{fig1}a)
\begin{equation}
  \frac{dN_{v}}{dt} = W^{\rm cap}_{v} (N_{v}+1) 
   - (W^{\rm down}_{v}+W^{\rm up}_{v})N_{v} \; ,
\end{equation}
where $W^{\rm cap}_{v}$ is the capture rate from the condensate,
and $W^{\rm down}_{v}$ and $W^{\rm up}_{v}$ are the loss rates
to more deeply bound states and back to the condensate,
respectively. The factor $N_{v}+1$ comes from the bosonic
nature of the atoms: the capture will be bose-enhanced (or 
stimulated), as opposed to the depletion of the level
(proportional to $N_{v}$).

Let us first calculate $W^{\rm cap}_{v}$. As mentioned above,
the capture rate is equal to the phonon emission rate.
For Bogoliobov quasi-particles (or phonons), the emission 
rate of phonons with momemtum $\vec{q}$ and energy $\hbar \omega_{\vec{q}}$ 
can be estimated using the Fermi Golden rule 
\cite{eddy,kittel} 
\begin{equation}
 w_{\rm emis.}(\vec{q}) = \frac{2\pi}{\hbar} \frac{\mu_{c}^{2}}{nV}
  \frac{\hbar^{2}q^{2}}{2m} \frac{(n_{\vec{q}}+1)}{\hbar \omega_{\vec{q}}}
 \delta (\Delta \varepsilon - \hbar \omega_{\vec{q}})
I(\vec{q}) \; ,
\end{equation} 
where $V$ is the volume occupied by the condensate, 
$\mu_{c}=4\pi\hbar^{2}na/m$ its chemical potential
($n$, $m$, and $a$ are the density, mass, and 
scattering length of the condensate atoms, respectively),
and $n_{\vec{q}}$ the phonon occupation number.
Here, $I(\vec{q})$ is the square of the form factor for
transitions from a continuum to a bound state.
If we represent the initial condensate state
by the single particle wavefunction $\Psi_{0}(\vec{r})$
and the final bound state by
the wavefunction $\Psi_{v}(\vec{r})$, with energy
difference 
$\Delta \varepsilon = \varepsilon_{0}-\varepsilon_{v}=\hbar \omega_{\vec{q}}$,
then
\begin{equation}
  I(\vec{q}) = N\left| \int d^{3}r \Psi^{*}_{v}(\vec{r})
   e^{-i\vec{q}\cdot\vec{r}} \Psi_{0}(\vec{r}) \right|^{2} \; ,
  \label{form-factor}
\end{equation}
where $N=nV$ is the number of atoms in the condensate.

At zero-temperature in an infinite homogeneous system, 
we set $\varepsilon_{0} =0$ ($\vec{k}=0$), 
and we have \cite{note-psi-correction}
\begin{equation}
 \Psi_{0}(\vec{r}) = \frac{1}{\sqrt{V}} \makebox[.5in]{and} 
 \Psi_{v}(\vec{r}) = \frac{1}{\sqrt{2\pi a_{v}}} \frac{e^{-r/a_{v}}}{r} \; ,
 \label{eq:wavefunctions}
\end{equation}  
where  
the binding energy of the final bound state  
is related to the extent of the $s$-wavefunction $\Psi_{v}(\vec{r})$
via $\varepsilon_{v} = -\hbar^{2}/2\mu a_{v}^{2}$,
with $\mu$ being the system reduced mass. Note that we have
$\Delta\varepsilon = \hbar^{2}/2\mu a_{v}^{2}$.
In an isotropic Bose gas, 
$I(\vec{q})$ depends only on the magnitude of $\vec{q}$, 
and with Eq.(\ref{eq:wavefunctions}),  
$I(\vec{q}) = 8\pi a_{v}^{3}n/(1+q^{2}a_{v}^{2})^{2}$.
              
\begin{figure}[h]
 \centerline{\epsfxsize=3.25in\epsfbox{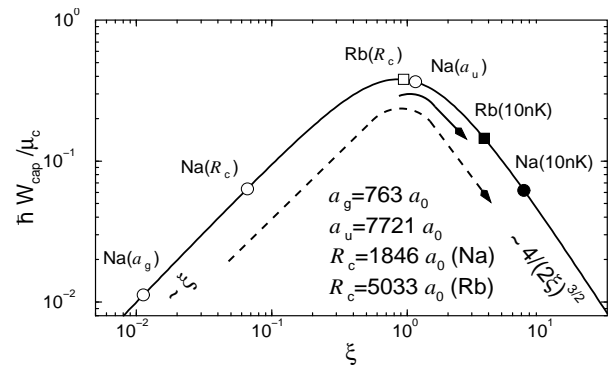}}
\caption{ \protect\narrowtext
          $W_{v}^{\rm cap}$ as a function of $\xi$ for various
          values of $a_{v}$ corresponding to the states 
          $^{2}\Sigma_{g}^{+}$, and $^{2}\Sigma_{u}^{+}$ of 
          Na$_{2}^{+}$ \protect\cite{hopping}, as well as $a_{v}\sim R_{c}$
          for both Na and Rb. The equilibrium points for Na and Rb
          at 10 nK are shown, as well as the two limiting behaviours.}
\label{fig2}
\end{figure}

The capture rate is obtained 
by integrating over all possible phonon states, i.e. 
$\int d^{3}q\; w_{\rm emis.}(\vec{q})\; V/(2\pi)^{3}$,
and
\begin{eqnarray}
 W^{\rm cap}_{v} =  4\frac{\mu_{c}}{\hbar} \sqrt{\frac{m}{\mu}} 
 \frac{a}{a_{v}} \frac{(\sqrt{1+\xi^{2}}-\xi)^{3/2}}
      {\sqrt{1+\xi^{2}}}
      (n_{q_{0}}+1)I(q_{0}) \; ,
 \label{rate-one}
\end{eqnarray}
where $\hbar \omega_{q_{0}}=\Delta\varepsilon$, and
$\xi =\mu_{c}/\Delta\varepsilon = 8\pi n a_{v}^{2}a \mu/m$. 
For Bogoliobov phonons, 
$\hbar \omega_{q} = \hbar q s \sqrt{1+(\hbar q/2ms)^{2}}$, and
we can express $q_{0}$ as 
\begin{equation}
  q_{0}^{2}a_{v}^{2} \frac{\mu}{m} = \sqrt{1+\xi^{2}}-\xi \; ,
  \label{eq:q0}
\end{equation}
where the sound velocity $s$ is related to the chemical potential
by $\mu_{c}=ms^{2}$, and $W^{\rm cap}_{v}$ becomes 
\begin{eqnarray}
  W^{\rm cap}_{v} & = & 4\frac{\mu_{c}}{\hbar} 
     \left( \frac{m}{\mu}\right)^{3/2} 
     \frac{\xi (\sqrt{1+\xi^{2}}-\xi)^{3/2}}
      {\sqrt{1+\xi^{2}}} \nonumber \\
   & & \times  \left[ 1 + \frac{m}{\mu}\left(\sqrt{1+\xi^{2}}-\xi\right)
     \right]^{-2} \makebox[-.15in]{}
      (n_{q_{0}}+1) \; .
      \label{eq:cap-rate}
\end{eqnarray}
For infinitly massive ions, we have $m/\mu =1$ \cite{note-mass}.
Note that the two limiting values of $\xi$ correspond to two 
different physical conditions: $\xi\rightarrow\infty$ implies
$q_{0}\rightarrow 0$, i.e. a phonon-like regime, and
$\xi\rightarrow 0$ implies dilute conditions with a
binary collision regime. 
In Fig.~\ref{fig2}, we illustrate $W^{\rm cap}_{v}$ as a function of $\xi$,
and show its behavior for the asymptotic regimes: 
$W^{\rm cap}_{v}\propto n^{2}a_{v}^{2}a^{2}$ for small $\xi$,
and $W^{\rm cap}_{v}\propto 1/a_{v}^{3}\sqrt{na}$ for large $\xi$.
The binary regime $(\xi\ll 1)$ is proportional
to $n^{2}$ as expected. The $n^{-1/2}$ dependence for $\xi \gg 1$
clearly indicates the dominant contribution of phonon assisted
transitions \cite{kittel,landau}. The increase in sound velocity
$s\propto \sqrt{n}$ with the condensate density $n$ reduces the rate of
phonon emission (and $W^{\rm cap}_{v}$) \cite{note-rate-phonon}.

As an example, we consider a sodium ion (Na$^{+}$) in a condensate of
sodium atoms at $T=0$, with $n\sim 10^{14}$ cm$^{-3}$ and 
$a=52$ $a_{0}$ \cite{ketterle} ($a_{0}$: bohr radius). 
For the uppermost bound state of the polarization potential, 
$a_{v}$ is well approximated by the atom-ion 
scattering length $a_{i}\sim 2000$ $a_{0}$ \cite{hopping,note-a}, 
and assuming $m/\mu \approx 1$ \cite{note-mass},
we obtain $\xi \sim 0.066$, and neglecting $n_{q_{0}}$, 
$W^{\rm cap}_{v}\sim 600$ s$^{-1}$ (see Fig.~\ref{fig2}).
Under these conditions, roughly 600 atoms will be captured
by the ion, and another 600 will be emitted as phonons (excited out
of the condensate) per second. In view of typical condensate lifetimes, 
this represents a sizable transfer mechanism that should be
observable. 

By comparison, the capture rate into more deeply bound states will
be much smaller. E.g., using the approach
of LeRoy and Bernstein \cite{leroy}, we find for the second 
uppermost level of the pure polarization potential
\begin{equation}
 a_{v-1}^{2} = \frac{a_{v}^{2}}{1+ (2\mu a_{v}^{2}/\hbar^{2})^{1/4}K_{4}}
\end{equation}
where 
$K_{4} = 4\sqrt{2\pi\hbar^{2}/\mu}\; \Gamma (5/4)/[\Gamma (3/4) C_{4}^{1/4}]$.
Because $a_{v-1}<a_{v}\sim a_{i}$, the capture rate will be accompanied by 
atom-like excitations with $W_{v}^{\rm cap}\propto a_{v}^{2}$,
and 
\begin{equation}
  \frac{W^{\rm cap}_{v}}{W^{\rm cap}_{v-1}} \approx 
  \frac{a_{v}^{2}}{a_{v-1}^{2}} \simeq 
  1+ \left( \frac{2\mu a_{v}^{2}}{\hbar^{2}}\right)^{1/4}K_{4}\; .
  \label{eq:capture-comp}
\end{equation}
For Na,  $C_{4}=162.7$ a.u. \cite{hopping}, and assuming
$m/\mu \approx 1$, 
we find that $W^{\rm cap}_{v}$ is about ten times
larger for the uppermost level than $W^{\rm cap}_{v-1}$
for the second uppermost level.
The ratio is even larger for deeper levels: 
spontaneous capture is important mostly in
the uppermost level.

Let us now examine the escape rates out of the bound states $v$,
$W^{\rm up}_{v}$ and $W^{\rm down}_{v}$. Since only the capture in
the uppermost level is relevant, we consider only this level.
At very low temperatures, $k_{B}T\ll \Delta \varepsilon$, and
there are no thermal phonons with sufficient energy to promote
bound atoms to the free condensate state, and non-equilibrium phonons 
emitted from the capture process are 
moving far away from the location of the ion and can be 
neglected. Hence, because of energy conservation, transitions
from bound levels to the continuum (into the condensate)
are not allowed, and $W_{v}^{\rm up}\sim 0$. 
The collision of atoms within the upper bound level (or condensate
atoms with the trapped ones) may result in the decay of the upper level
to a lower level. The rate of this process ($W^{\rm down}_{v}$) is  
inversely proportional to the binding energy of the deeper level: 
$W^{\rm cap}_{v}/W^{\rm down}_{v} \sim a_{v}^{2}/a_{v-1}^{2}$.
Hence, 
$W^{\rm down}_{v}$ is of the same order as $W^{\rm cap}_{v-1}$, 
and is at least one order of magnitude smaller than $W^{\rm cap}_{v}$.

As the number of trapped atoms $N_{v}$ within the uppermost bound 
level $v$ increases, the atom-atom
repulsive meanfield energy \cite{note:negative-a}
will grow, and will eventually ``push" the 
energy level up. Using a simple
treatment based of the Gross-Pitaevskii equation, one can show that
the binding energy of the uppermost level $v$, for
$N_{v}\gg 1$, is approximately given
by $\varepsilon_{v} (N_{v}) \simeq 
(ma_{v}/6\mu a N_{v})^{2/3}\varepsilon_{v}$.
Although the number of atoms trapped in the level $v$ 
would apparently grow to extremely large numbers (since the decay rates
out of it are much smaller), other effects,
such as charge hopping \cite{hopping} or thermal fluctuations, 
will limit that maximum number. 
The most important one is related to
thermal fluctuations. If their energy 
($\sim k_{B}T$) is equal to $|\varepsilon_{v} (N_{v})|$, 
thermal equilibrium is reached, and as many atoms get 
kicked out of the level as there are being captured. 
In other words, $W^{\rm up}_{v}$
will be equal to $W^{\rm cap}_{v}$.
At that point, $N_{v}^{\rm max}$ can be found from 
$\varepsilon_{v} (N_{v}^{\rm max})\sim k_{B}T$, and  
$\xi$ will have reached its maximal
value $\xi\sim\mu_{c}/k_{B}T$. 
For a BEC of Na atoms with $a_{v}\sim R_{c}\sim 2000 a_{0}$ 
and $T\sim$ 100 nK, 
we get $N_{v}^{\max}\sim 600$.
In Fig.~\ref{fig2}, we illustrate the trajectory of the system as a 
function of $\xi$ for various initial values of $a_{v}$, for both
sodium and rubidium condensates. In all cases, as $\xi$ grows 
from its initial to its final equilibrium
value, the capture rate $W_{v}^{\rm cap}$  passes
through a maximum before reaching its final value:
then the rates ``in" and ``out" of the uppermost level are equal, and
the system has reached thermal equilibrium.

Lasers can also be used to control  
coherent stimulated transitions of free condensate atoms
into a specific bound state $v$ of 
the molecular ion. Two off-resonant laser 
fields with a frequency difference 
matching the transition energy into a specific bound state $v$ cause a 
stimulated two-photon transition from the condensate into the bound state
of molecular ion (see Fig.~\ref{fig1}b). We assume that two 
laser beams are co-propagating, hence
negligible momentum is transfered to the atoms by the lasers. 
In the limit of 
sufficiently slow excitation, the atom-laser field interaction is described
by 
$H_{\rm int} = \hbar \Omega \int d^3 r \Psi_v(\vec{r})^* \Psi_{0}(\vec{r})
{\hat b}_v^\dagger{\hat b}_0  + h.c.,$
where $\Omega$ is the Rabi-frequency of the two-photon transition 
proportional to 
the laser intensity, and ${\hat b}_0,{\hat b}_v$ are bose annihilation 
operators for the condensate and bound atoms, respectively. 
The resulting dynamics 
corresponds approximately to that of a two-component condensate 
with effective coupling.
The number of atoms stimulated into the bound state can be estimated
by solving the equations of motion for ${\hat b}_0,{\hat b}_v$.
For a sufficently short interaction time $\tau$, assuming an undepleted
condensate $\langle{\hat b}_0^\dagger{\hat b}_0\rangle = N$, and neglecting 
spontaneous relaxation, we obtain for 
$N_v =\langle{\hat b}_v^\dagger{\hat b}_v\rangle$
\begin{equation}
N_v (\tau ) \simeq  |\Omega|^2 \tau^2 I_v(0)=8\pi a_{v}^{3} 
|\Omega|^2 \tau^2 N /V \; . 
\end{equation} 
The rate of stimulation $W_{v}^{\rm st}$
is therefore proportional to the laser power with a characteristic 
frequency scale given by $W_{v}^{\rm st}\sim (\Omega \sqrt{a_{v}^{3} N/V} )$, 
which can be easily made larger than spontaneous capture rate
$W_{v}^{\rm cap}$.

In practice, very fast excitation by lasers might result in heating of
the condensate. The heating rate $W_{v}^{\rm heat}$ is given by the 
transition probability from the bound state into continuum  (Fig.~\ref{fig1}c). 
This rate is also proportional to the laser power and can be 
estimated using Fermi's Golden rule. Taking  $1/W_{v}^{\rm st}$ as a  
characteristic transition time, we find that heating is negligible
in dilute condensate ($Na^3/V \ll 1$): $W_{v}^{\rm heat} \ll W_{v}^{\rm st}$
provided that 
\begin{equation}
W_{v}^{\rm st}\ll\hbar/m (N/V)^{2/3}. 
\end{equation} 
For realistic parameters, such as those of a sodium
condensate used above, the stimulated rate $W_{v}^{\rm st}$ can be in 
the range of
100KHz - 1 MHz
without substantial heating of the condensate.  
For weakly bound states, such rates
are achievable with modest laser powers used, e.g., in Bragg
spectroscopy experiments \cite{stenger}. It is important to note that 
as the number of bound atoms increases, atom-atom interaction
can shift the bound state energy out of resonance, thereby inhibiting 
coherent transitions. This is directly analogous to the recently proposed
``dipole blocade'' in Rydberg atoms \cite{dipole-blockade}, 
and can be potentially used for
non-trivial manipulation of quantum states in a BEC.

In conclusion, we predicted the possiblity of forming large molecular ions
in a BEC, analyzed the dynamics of their spontaneous growth and proposed
a technique for their fast and coherent generation in specific states. 
Such mesoscopic molecular ions could be observed, e.g.,
in experiments involving weak ionization of BECs. Of further interest
are experiments in which charged impurities are introduced 
via controlled processes involving either a combination of ion and atom 
traps, or using surface microtraps. The present work suggests that 
in such situations, ionic impurities can be used as a new tool for
the accurate manipulation of condensates, including their quantum state.
Finally we note that one could manipulate, using external electric and 
magnetic fields, the position of the atomic cloud trapped around
a charged particle.

The work of R.C. was supported by
the National Science Foundation through the Grant PHY-9970757.
The work of V.K. and M.L. was supported by 
National Science Foundation through ITAMP.
This work was also partially supported by MIT-Harvard Center for Ultracold
Atoms. 


 \end{multicols}


\begin{thebibliography}{99}
\bibitem{note-ion}
  The results obtained here are also valid for atoms and ions
  of different species (with negligible charge transfer).
\bibitem{note-recoil}
  The kinetic energy is shared between phonon and molecular ion: with
  its rapid growth, the molecular ion becomes very massive and
  recoil will be negligible.
\bibitem{note-psi-correction}
  Replacing $\Psi_{0}(\vec{r})$ by $[\cos (R_{c}/r) -
  (a_{i}/R_{c})\sin (R_{c}/r)]/\sqrt{V}$, where $a_{i}$
  is the atom-ion scattering length and 
  $R_{c}=\sqrt{\mu C_{4}}/\hbar$ is the long range scaling
  parameter, provides
  corrections for the presence of the ion.
  In the phonon regime ($\xi \ll 1$), this amounts  
  to multiply $W_{v}^{\rm cap}=4\mu_{c}/\hbar (2\xi )^{3/2}$ by  
  $(\pi R_{c}/4a_{i})^{2} + 1/9\xi^{2}$ (with
  $m/\mu =1$).
\bibitem{note-mass}
  The rapid capture of atoms will ensure the validity of the
  infinite ion mass approximation $m/\mu \approx 1$.
\bibitem{note-rate-phonon}
  This is analogous to the spontaneous photon emission rate 
  $W\sim 1/c^{3}$. Here,
  we have $W_{v}^{\rm cap}\sim \mu_{c}/s^{3}\sim 1/\sqrt{n}$.
\bibitem{note-a}
  The value of $a_{i}$ is of the order of $R_{c}$ 
  \cite{note-psi-correction}.
\bibitem{note:negative-a}
  If the scattering length is negative, the meanfield energy will make
  the bound level deeper instead.
\end{thebibliography}
\end{document}